\documentclass[epj]{svjour}
\usepackage{amsmath}
\usepackage{stmaryrd}
\usepackage{textcomp}
\usepackage{cite}
\usepackage{graphicx}
\usepackage{subfigure}
\usepackage{float}
\usepackage{lineno,hyperref}
\usepackage{lscape}
\usepackage[misc]{ifsym}
\usepackage{graphics}
\usepackage{enumerate}
\begin{document}
\title{Bianchi type-I Universe with Cosmological constant and quadratic equation of state in  $f(R,T)$ modified gravity }
\authorrunning{G. P. Singh \and Binaya K. Bishi}
\titlerunning{ B.V.S Cosmological model in $f(R,T)$ gravity }
\author{G. P. Singh\inst{1} \and Binaya K. Bishi\inst{2}% etc
%}                     % Do not remove
\thanks{\inst{1} \email{gpsingh@mth.vnit.ac.in} \\${}$\hspace{0.5cm} \inst{2} \email{binaybc@gmail.com}}}

%\thanks{\inst{2} Insert the address here if needed}
%}
\institute{Department of Mathematics, \\Visvesvaraya National Institute of Technology, \\Nagpur-440010, India}
\date{Received: date / Revised version: date}
\abstract{
This article deals with the study of Bianchi type-I universe in the context of $f(R,T)$ gravity. Einstein's field equations in $f(R,T)$ gravity has been solved in presence of cosmological constant $\Lambda$ and quadratic equation of state (EoS) $p=\alpha \rho^2-\rho$, where $\alpha\neq 0$ is a constant. Here we have discussed two classes of $f(R,T)$ gravity i.e. $f(R,T)=R+2f(T)$ and $f(R,T)=f_1(R)+f_2(T)$.  A set of  models has been taken into consideration based on  the plausible relation. Also we have studied the some physical and kinematical properties of the models.
 % end of PACS codes
} %end of abstract
\maketitle
\section{Introduction}
%\begin{sloppypar}
\label{introduction}
\par
It is known that \cite{Riess98,Perlmutter99,Percival01,Jimenez03,Stern10} in present scenario, our universe is accelerating. However, final satisfactory explanation about physical mechanism and driving force of accelerated expansion of the universe is yet to achieve as human mind has not achieved perfection.  From the modern cosmology, it is known that a point of universe is filled with dark energy. It has been addressed by various slow rolling scalar fields. It is supposed that the dark energy is responsible to produce sufficient acceleration in late time of evolution of the universe. Thus it is much more essential to study the fundamental nature of the dark energy and several approaches have been made to understand it. The cosmological constant is assumed to be the simplest candidate of dark energy. It is the classical correction made to the Einstein’s field equation by adding cosmological constant to the field equations. The introduction of cosmological constant to Einstein’s field equation is the most efficient way of generating accelerated expansion but it faces serious problems like fine-tuning and cosmic coincidence problem in cosmology \cite{Peebles03,Sahni00}. Quintessence \cite{Martin08}, phantom \cite{Nojiri06}, k-essence \cite{Chiba00}, tachyons \cite{Padmanabhan02}, Chaplygin gas \cite{Bento02} are the other representative of dark energy. However, there is no direct detection of such exotic fluids. Researchers are taking interest to explore dark energy due to lack of strong evidence of existence of dark energy. Several authors (Singh et al.\cite{SINGH06}, Singh et al.\cite{SINGH13}, Jamil and Debnath \cite{Jamil11}) have discussed cosmological model with cosmological constant in different contexts.
\par
Dark energy can be explored in several ways and modifying the geometric part of the Einstein-Hilbert action \cite{Magnano87} is treated as the most efficient possible way. Based on its modifications, several alternative theories of gravity came into existence. Some of the modified theories of gravity are $f(R)$, $f(T)$, $f(G)$ and $f(R,T)$ gravity. These models are proposed to explore the dark energy and other cosmological problems. Sharif and Azeem \cite{Sharif12} discussed the Cosmological evolution for Dark Energy Models in $f(T)$ Gravity. The $f(R)$ modified theory produces both cosmic inflation and mimic behavior of DE including present cosmic acceleration \cite{Felice05,Sotiriou10,Nojiri11}. Amendola et al.\cite{Amendola07} have discussed the cosmologically viable conditions in $f(R)$ theory which describe the dark energy models. Recently author like Sahoo and Sivakumar \cite{Sahoo15}, Ahmed and Pradhan \cite{Ahmed14}, Pradhan et al.\cite{Pradhan15} have investigated the cosmological models with cosmological constant in $f(R,T)$ gravity for different Bianchi type space-time.
\par
Quadratic equation of state is needed to explore in cosmological models due to its importance in brane world model and the study of dark energy and general relativistic dynamics for different models.  The general form of the quadratic equation of state is given by
\begin{equation}
\label{I1}
p=p_0+\alpha \rho +\beta \rho^2,
\end{equation}
where $p_0$,$\alpha$,$\beta$ are parameters. This equation \eqref{I1} is nothing but first term of Taylor expansion of any equation of state of the form $p=p(\rho)$ about $\rho=0$.
\par
Ananda and Bruni discussed the cosmological models by considering different form of non-linear quadratic equation of state. Ananda and Bruni \cite{Ananda05} have investigated the general relativistic dynamics of RW models with a  non-linear quadratic equation of state and analysed that the behaviour of the anisotropy at the singularity found in the brane scenario can be recreated in the general relativistic context by considering a equation of state of the form \eqref{I1}.  Again they \cite{Ananda06} have discussed the anisotropic homogeneous and inhomogeneous cosmological models in general relativity with the equation of state of the form
\begin{equation}
\label{I2}
p=\alpha \rho+\frac{\rho^2}{\rho_c},
\end{equation}
and tried to isotropize the universe at early times when the initial singularity is approached. In our present study, we have consider the quadratic equation of state of the form
\begin{equation}
\label{I3}
p=\alpha \rho^2-\rho ,
\end{equation}
where $\alpha \neq o$ is a constant quantity and such type of consideration does not effect the quadratic nature of equation of state.
\par
Nojiri and Odintsov \cite{Nojiri05} studied the effect of modification of general equation of state of dark energy ideal fluid by the insertion of inhomogeneous, Hubble parameter dependent term in the late-time universe. The quadratic equation of state may describe the dark energy or unified dark energy \cite{Nojiri05,Capozziello06}. Rahman et al. \cite{Rahman09} investigated the construction of an electromagnetic mass model using quadratic equation of state in the context of general theory of relativity. Feroze and Siddiqui \cite{Feroze11} studied the general situation of a compact relativistic body by taking a quadratic equation of state for the matter distribution. Maharaj and Takisa \cite{Maharaj13} have investigated the regular models with quadratic equation of state. They have considered static and spherically symmetric space-time with a charged matter distribution and found new exact solutions to the Einstein-Maxwell system of equations which are physically reasonable.
\par
A Cosmological model based on a quadratic equation of State unifying vacuum energy, radiation, and dark energy  has been discussed by  Chavanis \cite{Chavanis13a} and  also a  cosmological model describing the early inflation, the intermediate decelerating expansion, and the late accelerating expansion by a quadratic equation of state have been investigated by the same author \cite{Chavanis13b}. Strange Quark Star Model with Quadratic Equation of State has been investigated by Malaver \cite{Malaver14} and they have obtained a class of models with quadratic equation of state for the radial pressure that correspond to anisotropic compact sphere, where the gravitational potential $Z$ depends on an adjustable parameter $n$. Recently Reddy et al.\cite{Reddy15} has studied the Bianchi type-I cosmological model with quadratic equation of state in the context of general theory of relativity.
\par
Motivated by the aforesaid research, we have investigated the Bianchi type-I cosmological model in $f(R,T)$ gravity with quadratic equation of state and cosmological constant . Here we have discussed two classes of $f(R,T)$ gravity.
\section{Gravitational field equations of $f(R,T)$ modified gravity theory}
Let us consider the action for the modified gravity as
\begin{equation}
\label{e1}
 S=\int \left(\frac{f(R,T)}{16\pi G}+L_m\right)\sqrt{-g}d^4x,
\end{equation}
where $f(R,T)$ is the arbitrary function of $R$ and $T$. $R$ is the Ricci scalar and $T$ is the tress of the stress-energy tensor of the matter $T_{ij}$. $L_m$ is the matter lagrangian density. For the choice of $f(R,T)$ we will get the action for the different theories . If $f(R,T)\equiv f(R)$ and $f(R,T)\equiv R$ then \eqref{e1} represents the action for $f(R)$ gravity and general relativity respectively. The stress energy tensor of matter is defined as
\begin{equation}
\label{e2}
T_{ij}=-\frac{2}{\sqrt{-g}}\frac{\delta(\sqrt{-g}L_m)}{\delta g^{ij}},
\end{equation}
and its stress by $T=g^{ij}T_{ij}$. If we consider that the matter lagrangian density $L_m$ of matter is depends only on $g_{ij}$ and not on its derivatives leads us
\begin{equation}
\label{e3}
T_{ij}=g_{ij}L_m-2\frac{\partial L_m}{\partial g^{ij}}
\end{equation}
By varying the action \eqref{e1} w.r.t the metric tensor component $g_{ij}$ we have
\begin{equation}
\label{e4}
f_R(R,T)R_{ij}-\frac{1}{2}f(R,T)g_{ij}+(g_{ij}\qed-\nabla_i\nabla_j)f_R(R,T)=8\pi-f_T(R,T)T_{ij}-f_T(R,T)\Theta_{ij},
\end{equation}
where
\begin{equation}
\label{e5}
\Theta_{ij}=-2T_{ij}+g_{ij}L_m-2g^{lk}\frac{\partial^2 L_m}{\partial g^{ij}\partial g^{lk}}
\end{equation}
Here $f_T(R,T)=\frac{\partial f(R,T)}{\partial T}$,$f_R(R,T)=\frac{\partial f(R,T)}{\partial R}$, $\qed\equiv\nabla^i\nabla_j$ is the De Alembert's operator and $T_{ij}$ is the standard matter energy momentum tensor derived from the lagrangian $L_m$. By contracting equation \eqref{e4}, we obtained the relation between $R$ and $T$ as
\begin{equation}
\label{e6}
f_R(R,T)R+3\qed f_R(R,T)-2f(R,T)=8\pi T-f_T(R,T)T-f_T(R,T)\Theta,
\end{equation}
where $\Theta=g^{ij}\Theta_{ij}$. From \eqref{e4} and \eqref{e6}, the gravitational field equations can be written as
\begin{equation}
\label{e7}
f_R(R,T)(R_{ij}-\frac{1}{3}Rg_{ij})+\frac{1}{6}f(R,T)g_{ij}=(8\pi-f_T(R,T))(T_{ij}-\frac{1}{3}Tg_{ij})-f_T(R,T)(\Theta_{ij}-\frac{1}{3}\Theta g_{ij})+\nabla_i\nabla_j f_R(R,T)
\end{equation}
The perfect fluid form of the stress energy tensor of the matter lagrangian is given by
\begin{equation}
\label{e8}
T_{ij}=(\rho+p)u_iu_j-pg_{ij},
\end{equation}
where $u^i=(1,0,0,0)$ is the four-velocity vector and satisfies the relation $u^iu_i=1$ and $u^i\nabla_ju_i=0$. $\rho$ and $p$ are the energy density and pressure of the fluid respectively. From equation \eqref{e5} we have
\begin{equation}
\label{e9}
\Theta_{ij}=-2T_{ij}-pg_{ij}
\end{equation}
It is to note that the functional $f(R,T)$ is depends on the physical nature of the matter field through the tensor $\Theta_{ij}$. Thus for each choice of $f(R,T)$ leads us to different cosmological models. Herko et al \cite{Harko11} presented three class of $f(R,T)$ as follows
\begin{equation}
\label{e10}
    f(R,T)=
\begin{cases}
    R+2f(T)\\
    f_1(R)+f_2(T)\\
    f_1(R)+f_2(R)f_3(T)
\end{cases}
\end{equation}
In this present work we have discussed two class of $f(R,T)$ i.e. $f(R,T)=R+2f(T)$ and $f(R,T)=f_1(R)+f_2(T)$.\\
For the choice of $f(R,T)=R+2f(T)$ and with the help of \eqref{e8} and \eqref{e9}, equation \eqref{e4} takes the form
\begin{equation}
\label{e11}
G_{ij}=(8\pi+2f'(T))T_{ij}+(2pf'(T)+f(T))g_{ij}.
\end{equation}
Which is the gravitational field equation in $f(R,T)$ modify gravity for the class $f(R,T)=R+2f(T)$.
For the choice of $f(R,T)=f_1(R)+f_2(T)$ and with the help of \eqref{e8} and \eqref{e9}, equation \eqref{e4} takes the form
\begin{equation}
\label{e12}
f_1'(R)R_{ij}-\frac{1}{2}f_1(R)g_{ij}+(g_{ij}\qed-\nabla_i\nabla_j)f_1'(R)=(8\pi+f_2'(T))T_{ij}+(f_2'(T)p+\frac{1}{2}f_2(T))g_{ij}
\end{equation}
Which is regarded as the gravitational field equation in $f(R,T)$ modify gravity for the class $f(R,T)=f_1(R)+f_2(T)$.
\section{Field equations and cosmological model for $f(R,T) = R +2f(T)$}
In  $f(R,T)$ theory, the gravitational field equations  \eqref{e11} in presence of Cosmological constant $\Lambda$ are given as \begin{equation}
\label{e13}
 G_{ij}=\left[8\pi+2f'(T)\right]T_{ij}+\left[2pf'(T)+f(T)+\Lambda\right]g_{ij},
\end{equation}
where  prime denotes differentiation with respect to the argument.
For the choice of $f(T)=\lambda T$ ,eqn.\eqref{e13} takes the form
\begin{equation}
\label{e14}
 G_{ij}=\left[8\pi+2\lambda)\right]T_{ij}+\left[\lambda \rho-p\lambda+\Lambda\right]g_{ij}
\end{equation}
Let us consider the  Bianchi type-I space-time in the form
\begin{equation}
\label{e15}
ds^2=dt^2-X_1^2dx^2-X_2^2dy^2-X_3^2dz^2,
\end{equation}
where $X_1$, $X_2$ and $X_3$ are function of $t$ only. The field equations\eqref{e14} for the line element \eqref{e15} takes the form
\begin{equation}
\label{e16}
\frac{\dot{X_1}\dot{X_2}}{X_1X_2}+\frac{\dot{X_1}\dot{X_3}}{X_1X_3}+\frac{\dot{X_2}\dot{X_3}}{X_2X_3}=-(8\pi+3\lambda)\rho+p\lambda-\Lambda
\end{equation}
\begin{equation}
\label{e17}
\frac{\ddot{X_2}}{X_2}+\frac{\ddot{X_3}}{X_3}+\frac{\dot{X_2}\dot{X_3}}{X_2X_3}=(8\pi+3\lambda)p-\lambda \rho-\Lambda
\end{equation}
\begin{equation}
\label{e18}
\frac{\ddot{X_1}}{X_1}+\frac{\ddot{X_3}}{X_3}+\frac{\dot{X_1}\dot{X_3}}{X_1X_3}=(8\pi+3\lambda)p-\lambda \rho-\Lambda
\end{equation}
\begin{equation}
\label{e19}
\frac{\ddot{X_1}}{X_1}+\frac{\ddot{X_2}}{X_2}+\frac{\dot{X_1}\dot{X_2}}{X_1X_2}=(8\pi+3\lambda)p-\lambda \rho-\Lambda
\end{equation}
\section{Solution Procedure}
Now our problem is to solve the Einstein's modified field equations \eqref{e16}-\eqref{e19}. Here the system having four equations and six unknowns ($X_1$, $X_2$, $X_3$, $p$, $\rho$ and $\Lambda$). To obtain the complete solution, we need two more physically plausible relations. The considered two physically plausible relations are
\begin{enumerate}
  \item Quadratic equation of state
  \item Expansion Law\begin{itemize}
                       \item Power law  \begin{equation}
                                                        \label{e20}
                                                        V=V_0t^{3n},
                                                      \end{equation}
                       \item Exponential law\begin{equation}
                                                       \label{e21}
                                                        V=V_0e^{\beta t},
                                                      \end{equation}
                     \end{itemize}
\end{enumerate}
where $n$ and $\beta$ are poitive constant quantity. According to the choice of expansion Law,we have obtained two different models of the Bianchi type-I universe.
\subsection{Power law Model}
\label{sub1}
With the help of equations \eqref{e17}-\eqref{e19},we have obtained the metric potentials as
\begin{equation}
\label{e22}
X_i(t)=X_{i0}V^{\frac{1}{3}}exp\left[\int\frac{X_{0i}}{V}\right],\: i=1,2,3 ,
\end{equation}
where $X_{i0}$ and $X_{0i}$ are constant of integration $(i=1,2,3)$  and which satisfies the relation $\prod_{i=1}^3X_{i0}=1$ and $\sum_{i=1}^3X_{0i}=0$.
From the equations \eqref{e16}-\eqref{e17} and along with \eqref{I3} we have got
\begin{equation}
\label{e23}
\rho^2=\frac{1}{\alpha(8\pi+2\lambda)}\left[\frac{\ddot{X_2}}{X_2}+\frac{\ddot{X_3}}{X_3}+\frac{\dot{X_2}\dot{X_3}}{X_2X_3}-\frac{\dot{X_1}\dot{X_2}}{X_1X_2}-\frac{\dot{X_1}\dot{X_3}}{X_1X_3}-\frac{\dot{X_2}\dot{X_3}}{X_2X_3}\right]
\end{equation}
Using \eqref{e20} in \eqref{e22}, we have the metric potential as
\begin{equation}
\label{e24}
X_i(t)=X_{i0}V^{\frac{1}{3}}exp\left[\frac{-X_{0i}t^{-3n+1}}{(3n-1)V_0}\right],\: i=1,2,3
\end{equation}
The directional Hubble parameters are obtained as $H_i=\frac{n}{t}+\frac{X_{oi}}{V_0t^{3n}}$, $i=1,2,3$. The Hubble parameter $(H)$, deceleration parameter $(q)$ , expansion scalar $(\Theta)$ and Shear scalar $(\sigma^2)$ are as follows
$$ H=\frac{n}{t}, \: q=-1+\frac{1}{n} ,\: \Theta=3\frac{n}{t},\: \sigma^2=\frac{X_{02}^2+X_{03}^2+X_{02}X_{03}}{V_0^2t^{6n}}$$
Using the observational value for $q = -0.33\pm 0.17 $ \cite{Kotambkar14}, we have restricted $n$ as $n \in (1.19,2)$
in case of Power law model. Here we noticed that $H$,$\Theta$ and $\sigma^2$ die out for larger values of $t$.
With the help of \eqref{e24} from \eqref{e23}, the energy density is obtained as
\begin{equation}
\label{e25}
\rho^2=\frac{1}{(4\pi+\lambda)\alpha}\left[\frac{X_{02}^2+X_{03}^2+X_{02}X_{03}}{V_0^2}\frac{1}{t^{6n}}-\frac{n}{t^2}\right]
\end{equation}
Using \eqref{e25} in \eqref{I3}, we have the pressure as
\begin{equation}
\label{e26}
p=\frac{(X_{02}^2+X_{03}^2+X_{02}X_{03})t^{-6n+2}-V_0^2n}{(4\pi+\lambda)V_0^2t^2}-\sqrt{\frac{(X_{02}^2+X_{03}^2+X_{02}X_{03})t^{-6n+2}-V_0^2n}{(4\pi+\lambda)V_0^2\alpha t^2}}
\end{equation}
With the help of \eqref{e24}-\eqref{e26} from \eqref{e16}, the cosmological constant $\Lambda$ is obtained as
\begin{eqnarray}
\label{e27}
\Lambda&=&\frac{-4}{(4\pi+\lambda)V_0^2t^{6n+2}} \bigg[(2\pi+\lambda)(4\pi+\lambda)V_0^2t^{6n+2}-\sqrt{\frac{(X_{02}^2+X_{03}^2+X_{02}X_{03})t^{-6n+2}-V_0^2n}{(4\pi+\lambda)V_0^2\alpha t^2}}\nonumber\\
&-& (\pi+\frac{\lambda}{2})(X_{02}^2+X_{03}^2+X_{02}X_{03})t^2+\frac{1}{4}(3\lambda n+\lambda+12n\pi)V_0^2nt^{6n}\bigg]
\end{eqnarray}
\begin{figure}[H]
\centering
\begin{minipage}[b]{0.35\linewidth}
\resizebox{3in}{2in}{
\includegraphics{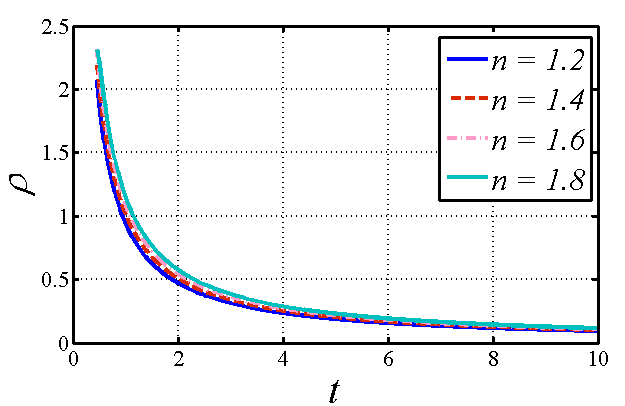}
}
\caption{Variation of energy density $\rho$ against time $t$ for $\lambda=1$ , $\alpha=-0.1$ ,$V_0=1$,$X_{02}=0.01$,$X_{03}=0.01$  and different $n(1.2,1.4,1.6,1.8)$ }.
\label{fig1}
\end{minipage}
\hspace{0.08\linewidth}
%\end{figure}
%\begin{figure}[H]
\centering
\begin{minipage}[b]{0.35\linewidth}
\resizebox{3in}{2in}{
\includegraphics{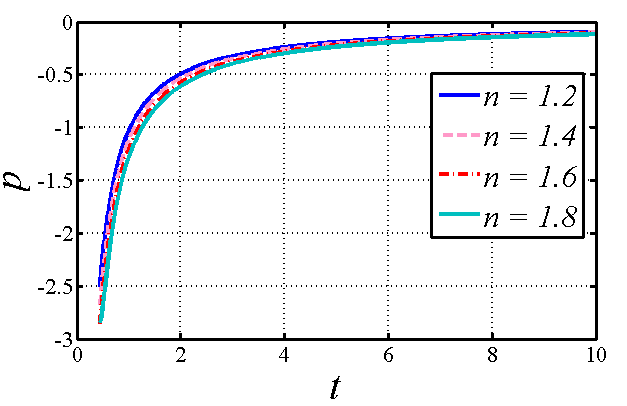}
}
\caption{Variation of Pressure $p$ against time $t$ for $\lambda=1$ , $\alpha=-0.1$ ,$V_0=1$,$X_{02}=0.01$,$X_{03}=0.01$  and different $n(1.2,1.4,1.6,1.8)$ }.
\label{fig2}
\end{minipage}
\end{figure}
\begin{figure}[H]
\centering
\resizebox{3in}{2in}{
\includegraphics{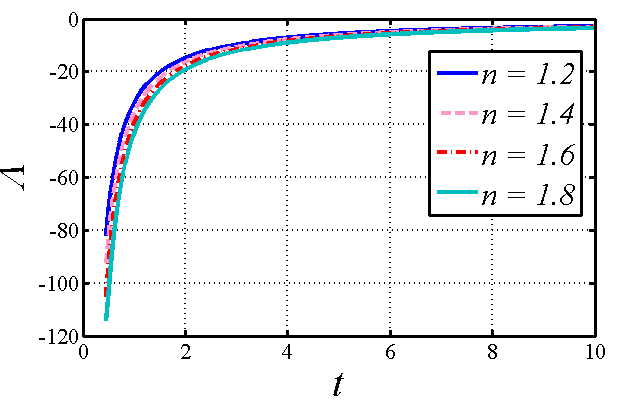}
}
\caption{Variation of Cosmological constant $\Lambda$ against time $t$ for $\lambda=1$ , $\alpha=-0.1$ ,$V_0=1$,$X_{02}=0.01$,$X_{03}=0.01$  and different $n(1.2,1.4,1.6,1.8)$ }.
\label{fig3}
\end{figure}
Fig.\ref{fig1} and Fig.\ref{fig2} show the variation of energy density $\rho$ and pressure $p$ against time $t$  for different values as in the figures. Here we noticed that $\rho, p\rightarrow0$ when $t\rightarrow\infty$. In the increase of $n$, energy density and pressure increases and decreases respectively. Fig.\ref{fig3} represents the variation of cosmological constant $\Lambda$ against time $t$  for different values as in the figures. It is observed that Cosmological constant $\Lambda$ is decreasing with the increase of $n$ and with the evolution of time it approaches towards zero.
\subsection{Exponential  law Model}
\label{sub2}
In this case, with the help of \eqref{e21} in \eqref{e22},  we have found the metric potential as
\begin{equation}
\label{e28}
X_i(t)=X_{i0}V^{\frac{1}{3}}exp\left[-\left(\frac{-\beta^2V_0t+3X_{0i}e^{-\beta t}}{3\beta V_0}\right)\right],\: i=1,2,3
\end{equation}
The directional Hubble parameters are obtained as $H_i=\frac{\beta}{3}+\frac{X_{oi}}{V_0e^{\beta t}}$, $i=1,2,3$. The Hubble parameter $(H)$, deceleration parameter $(q)$ , expansion scalar $(\Theta)$ and Shear scalar $(\sigma^2)$ are as follows
$$ H=\frac{\beta}{3}, \: q=-1 ,\: \Theta=\beta,\: \sigma^2=\frac{X_{02}^2+X_{03}^2+X_{02}X_{03}}{V_0^2e^{2\beta t}}$$ Here we noticed that $\sigma^2$ die out for larger values of $t$.
From \eqref{e28}and  \eqref{e23}, the energy density is expressed as
\begin{equation}
\label{e29}
\rho^2=\frac{(X_{02}^2+X_{03}^2+X_{02}X_{03})e^{-2\beta t}}{V_0^2\alpha (4\pi+\lambda)}
\end{equation}
Using \eqref{e29} in \eqref{I3}, the pressure is expressed as
\begin{equation}
\label{e30}
p=\frac{1}{(4\pi+\lambda)V_0^2}\left[-(4\pi+\lambda)V_0^2\sqrt{\frac{(X_{02}^2+X_{03}^2+X_{02}X_{03})e^{-2\beta t}}{V_0^2\alpha (4\pi+\lambda)}}+(X_{02}^2+X_{03}^2+X_{02}X_{03})e^{-2\beta t}\right]
\end{equation}
With the help of \eqref{e28}-\eqref{e30} from \eqref{e16}, the cosmological constant $\Lambda$ is obtained as
\begin{equation}
\label{e31}
\Lambda=-(8\pi+4\lambda)\sqrt{\frac{(X_{02}^2+X_{03}^2+X_{02}X_{03})e^{-2\beta t}}{V_0^2\alpha (4\pi+\lambda)}}+\frac{2(2\pi+\lambda)(X_{02}^2+X_{03}^2+X_{02}X_{03})}{e^{2\beta t}V_0^2 (4\pi+\lambda)}-\frac{\beta^2}{3}
\end{equation}
\begin{figure}[H]
\centering
\begin{minipage}[b]{0.35\linewidth}
\resizebox{3in}{2in}{
\includegraphics{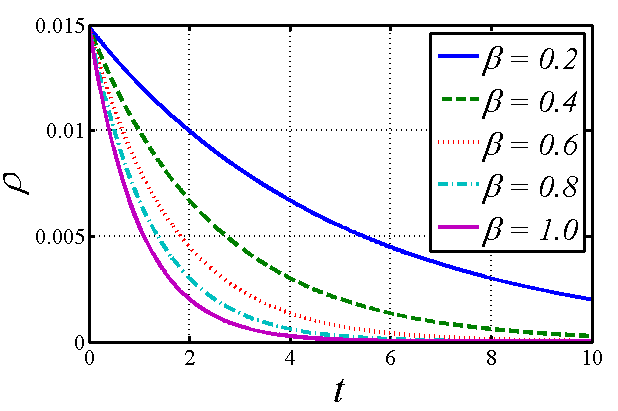}
}
\caption{Variation of energy density $\rho$ against time $t$ for $\lambda=1$ , $\alpha=0.1$ ,$V_0=1$,$X_{02}=0.01$,$X_{03}=0.01$  and different $\beta(0.2,0.4,0.6,0.8,1)$  }.
\label{fig4}
\end{minipage}
\hspace{0.08\linewidth}
%\end{figure}
%\begin{figure}[H]
\centering
\begin{minipage}[b]{0.35\linewidth}
\resizebox{3in}{2in}{
\includegraphics{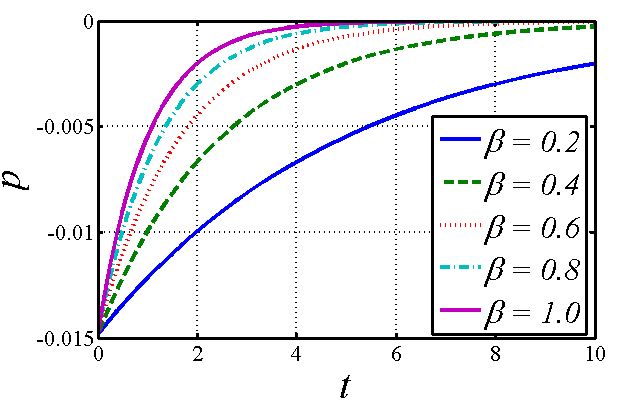}
}
\caption{Variation of Pressure $p$ against time $t$ for $\lambda=1$ , $\alpha=0.1$ ,$V_0=1$,$X_{02}=0.01$,$X_{03}=0.01$  and different $\beta(0.2,0.4,0.6,0.8,1)$ }.
\label{fig5}
\end{minipage}
\end{figure}
\begin{figure}[H]
\centering
\resizebox{3in}{2in}{
\includegraphics{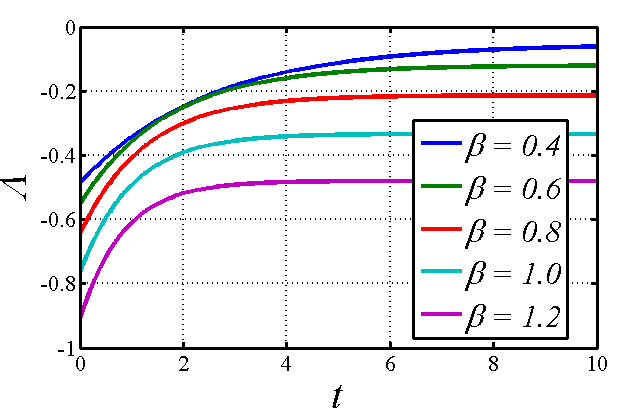}
}
\caption{Variation of Cosmological constant $\Lambda$ against time $t$ for $\lambda=1$ , $\alpha=0.1$ ,$V_0=1$,$X_{02}=0.01$,$X_{03}=0.01$  and different $\beta(0.4,0.6,0.8,1,1.2)$  }.
\label{fig6}
\end{figure}
Fig.\ref{fig4} and Fig.\ref{fig5} show the variation of energy density $\rho$ and pressure $p$ against time $t$  for different values as in the figures. Here we noticed that $\rho, p\rightarrow0$ when $t\rightarrow\infty$. In the increase of $\beta$, energy density and pressure decreases respectively. Fig.\ref{fig6} represents the variation of cosmological constant $\Lambda$ against time $t$  for different values as in the figures. It is observed that Cosmological constant $\Lambda$ is not approaching towards zero with the evolution of time and also it takes  negative values.
\section{Field equations and cosmological model for $f(R,T) = f_1(R)+f_2(T)$}
In  $f(R,T)$ theory, the gravitational field equations \eqref{e12} for the choice of $f_1(R)=\lambda R$ and $f_2(T)=\lambda T$, along with Cosmological constant $\Lambda$ are given as
\begin{equation}
\label{e32}
 G_{ij}=\left(\frac{8\pi+\lambda}{\lambda}\right)T_{ij}+\left(\frac{\rho-p+2\Lambda}{2}\right)g_{ij},
\end{equation}
In this case, the field equations are given by
\begin{equation}
\label{e33}
\frac{\dot{X_1}\dot{X_2}}{X_1X_2}+\frac{\dot{X_1}\dot{X_3}}{X_1X_3}+\frac{\dot{X_2}\dot{X_3}}{X_2X_3}=-\left(\frac{16\pi+3\lambda}{2\lambda}\right)\rho+\frac{p}{2}-\Lambda
\end{equation}
\begin{equation}
\label{e34}
\frac{\ddot{X_2}}{X_2}+\frac{\ddot{X_3}}{X_3}+\frac{\dot{X_2}\dot{X_3}}{X_2X_3}= \left(\frac{16\pi+3\lambda}{2\lambda}\right)p-\frac{\rho}{2}-\Lambda
\end{equation}
\begin{equation}
\label{e35}
\frac{\ddot{X_1}}{X_1}+\frac{\ddot{X_3}}{X_3}+\frac{\dot{X_1}\dot{X_3}}{X_1X_3}=\left(\frac{16\pi+3\lambda}{2\lambda}\right)p-\frac{\rho}{2}-\Lambda
\end{equation}
\begin{equation}
\label{e36}
\frac{\ddot{X_1}}{X_1}+\frac{\ddot{X_2}}{X_2}+\frac{\dot{X_1}\dot{X_2}}{X_1X_2}=\left(\frac{16\pi+3\lambda}{2\lambda}\right)p-\frac{\rho}{2}-\Lambda
\end{equation}
\subsection{Power Law Model}
Following the same procedure as in subsection \eqref{sub1}, we have obtained the same metric potential as in equation \eqref{e24} and the other parameters like energy density $\rho$, pressure $p$ and cosmological constant $\Lambda$ are expressed as follows
\begin{equation}
\label{e37}
\rho^2=\frac{2\lambda}{V_0^2\alpha (8\pi+\lambda)}
\left[\frac{X_{03}^2+X_{02}^2+X_{02}X_{03}}{t^{6n}}-\frac{V_0^2n}{t^2}\right]
\end{equation}
\begin{equation}
\label{e38}
p=\frac{2\lambda}{V_0^2(8\pi+\lambda)}
\left[\frac{X_{03}^2+X_{02}^2+X_{02}X_{03}}{t^{6n}}-\frac{V_0^2n}{t^2}\right]-\sqrt{\frac{2\lambda}{V_0^2\alpha (8\pi+\lambda)}
\left[\frac{X_{03}^2+X_{02}^2+X_{02}X_{03}}{t^{6n}}-\frac{V_0^2n}{t^2}\right]}
\end{equation}
\begin{eqnarray}
\label{e39}
\Lambda&=&\frac{1}{2t^2\lambda (8\pi+\lambda)\alpha V_0^2}\bigg[-41\sqrt{\lambda}(4\pi+\lambda)\sqrt{8\pi+\lambda}\sqrt{\alpha}V_0t^{-3n+1}\sqrt{2V_0^2nt^{6n}-2t^2(X_{03}^2+X_{02}^2+X_{02}X_{03})}\nonumber\\
&+&4(4\pi+\lambda)(X_{03}^2+X_{02}^2+X_{02}X_{03})\lambda \alpha t^{-6n+2}-2n(3\lambda n+\lambda+24n\pi)V_0^2\alpha \lambda\bigg]
\end{eqnarray}
\begin{figure}[H]
\centering
\begin{minipage}[b]{0.35\linewidth}
\resizebox{3in}{2in}{
\includegraphics{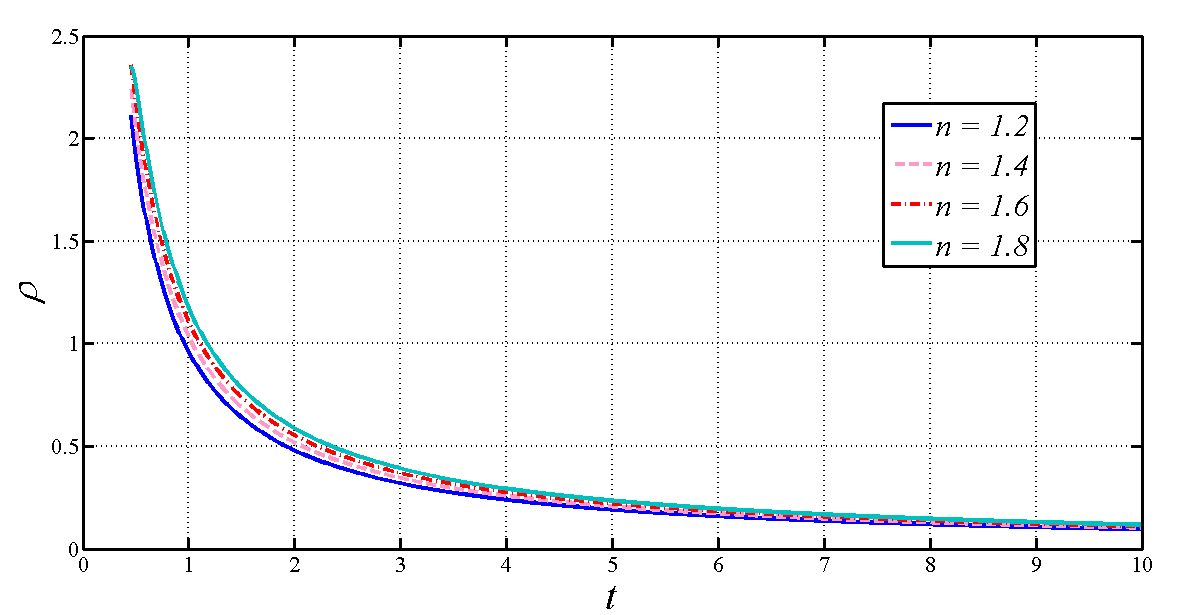}
}
\caption{Variation of energy density $\rho$ against time $t$ for $\lambda=1$ , $\alpha=-0.1$ ,$V_0=1$,$X_{02}=0.01$,$X_{03}=0.01$  and different $n(1.2,1.4,1.6,1.8)$ }.
\label{fig7}
\end{minipage}
\hspace{0.08\linewidth}
%\end{figure}
%\begin{figure}[H]
\centering
\begin{minipage}[b]{0.35\linewidth}
\resizebox{3in}{2in}{
\includegraphics{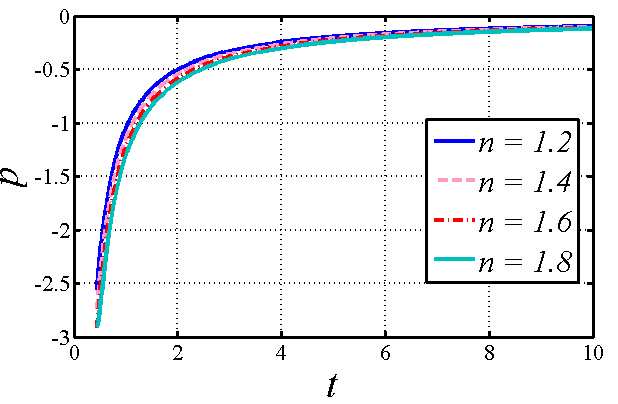}
}
\caption{Variation of Pressure $p$ against time $t$ for $\lambda=1$ , $\alpha=-0.1$ ,$V_0=1$,$X_{02}=0.01$,$X_{03}=0.01$  and different $n(1.2,1.4,1.6,1.8)$ }.
\label{fig8}
\end{minipage}
\end{figure}
\begin{figure}[H]
\centering
\resizebox{3in}{2in}{
\includegraphics{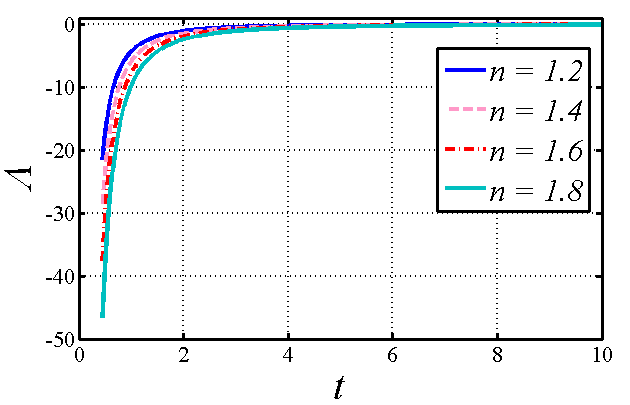}
}
\caption{Variation of Cosmological constant $\Lambda$ against time $t$ for $\lambda=1$ , $\alpha=-0.1$ ,$V_0=1$,$X_{02}=0.01$,$X_{03}=0.01$  and different $n(1.2,1.4,1.6,1.8)$ }.
\label{fig9}
\end{figure}
Here also we have noticed the similar results as in \eqref{sub1}.
\subsection{Exponential law Model}
Following the same procedure as in subsection \eqref{sub2}, we have obtained the same metric potential as in equation \eqref{e29} and the other parameters like energy density $\rho$, pressure $p$ and cosmological constant $\Lambda$ are as follows
\begin{equation}
\label{e40}
\rho=\frac{1}{V_0e^{\beta t}}\sqrt{\frac{2(X_{03}^2+X_{02}^2+X_{02}X_{03})\lambda}{\alpha (8\pi+\lambda)}}
\end{equation}
\begin{equation}
\label{e41}
p=\frac{2(X_{03}^2+X_{02}^2+X_{02}X_{03})\lambda}{V_0^2e^{2\beta t} (8\pi+\lambda)}-\frac{1}{V_0e^{\beta t}}\sqrt{\frac{2(X_{03}^2+X_{02}^2+X_{02}X_{03})\lambda}{\alpha (8\pi+\lambda)}}
\end{equation}
\begin{equation}
\label{e42}
\Lambda=-\frac{2\sqrt{2}(4\pi+\lambda)}{\lambda}\sqrt{\frac{2(X_{03}^2+X_{02}^2+X_{02}X_{03})\lambda e^{-2\beta t}}{V_0^2\alpha (8\pi+\lambda)}}+\frac{2(X_{03}^2+X_{02}^2+X_{02}X_{03})(4\pi+\lambda) e^{-2\beta t}}{V_0^2(8\pi+\lambda)}-\frac{\beta^2}{3}
\end{equation}
\begin{figure}[H]
\centering
\begin{minipage}[b]{0.35\linewidth}
\resizebox{3in}{2in}{
\includegraphics{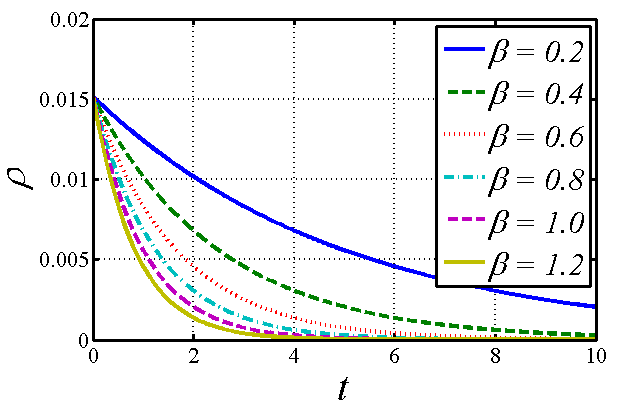}
}
\caption{Variation of energy density $\rho$ against time $t$ for $\lambda=1$ , $\alpha=0.1$ ,$V_0=1$,$X_{02}=0.01$,$X_{03}=0.01$  and different $\beta(0.2,0.4,0.6,0.8,1,1.2)$  }.
\label{fig10}
\end{minipage}
\hspace{0.08\linewidth}
%\end{figure}
%\begin{figure}[H]
\centering
\begin{minipage}[b]{0.35\linewidth}
\resizebox{3in}{2in}{
\includegraphics{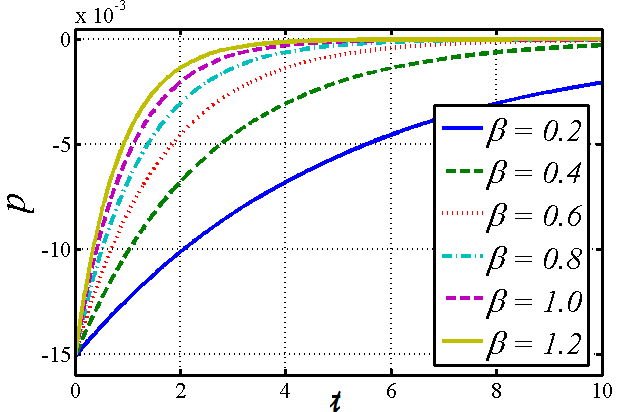}
}
\caption{Variation of Pressure $p$ against time $t$ for $\lambda=1$ , $\alpha=0.1$ ,$V_0=1$,$X_{02}=0.01$,$X_{03}=0.01$  and different $\beta(0.2,0.4,0.6,0.8,1,1.2)$ }.
\label{fig11}
\end{minipage}
\end{figure}
\begin{figure}[H]
\centering
\resizebox{3in}{2in}{
\includegraphics{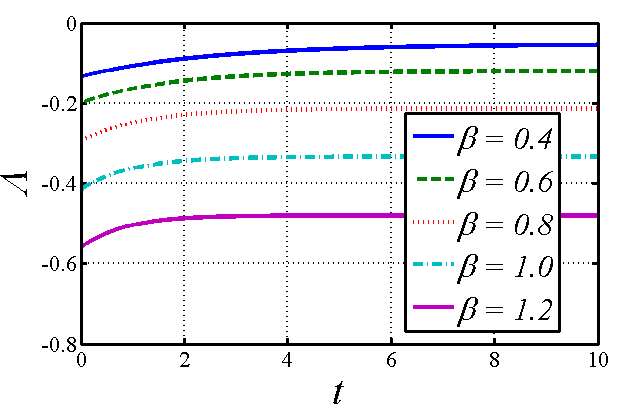}
}
\caption{Variation of Cosmological constant $\Lambda$ against time $t$ for $\lambda=1$ , $\alpha=0.1$ ,$V_0=1$,$X_{02}=0.01$,$X_{03}=0.01$  and different $\beta(0.4,0.6,0.8,1,1.2)$}.
\label{fig12}
\end{figure}
Here also we have noticed the similar results as in \eqref{sub2}.
\section{Concluding Remarks}
In this article we have the Bianchi type-I cosmological model in $f(R,T)$ modified gravity for two different classes of $f(R,T)$ in presence of cosmological constant and quadratic equation of state. Here we have discussed two models based the expansion law. From both the models, case of $f(R,T) = R +2f(T)$, we have concluded the following points.
\begin{itemize}
  \item In both the models, energy density $\rho$ is decreasing function of $t$ and $\rho$ approaches towards zero with the evolution of time.
  \item In both the models, pressure $p$ is negative and  approaches towards zero with the evolution of time.
  \item In both the models, cosmological constant $\Lambda$ is negative but here we notice that incase of Power law $\Lambda$ approaches towards zero with the evolution of time where as it does not approaches towards zero with the evolution of time incase of exponential law.
\end{itemize}
Similar observation also noticed for the case of $f(R,T) = f_1(R)+f_2(T)$. Here all the observation are in fare agreement with the observational data.

\end{document}